\documentclass[prb,preprint]{revtex4}
\usepackage [dvips]{graphicx}
\usepackage{amssymb}

\begin{document}

\title
{A NOVEL BROADBAND MEASUREMENT METHOD FOR THE MAGNETOIMPEDANCE OF RIBBONS AND THIN FILMS}

\author{A. Fessant, J. Gieraltowski, C. Tannous and R. Valenzuela \P}
\affiliation{UBO, LMB-CNRS UMR 6135, BP: 809 Brest CEDEX, 29285 France \\
\P Instituto de Investigaciones en Materiales, Universidad 
Nacional Aut\'{o}noma de Mexico, P.O. Box 70-360, Coyoacan, Mexico D.F., 04510, Mexico}

\begin{abstract}
A novel broad-band measurement method of the MI in thin films and ribbons is 
presented. It is based on the automated measurement of the reflection 
coefficient of a cell loaded with the sample. Illustrative results obtained 
with a permalloy multilayer thin film are presented and discussed.
\end{abstract}


\maketitle

\section{Introduction}

The magnetoimpedance (MI) effect consists of a change in the complex 
impedance of a ferromagnetic conductor under application of a static 
magnetic field $H_{dc}$. This effect has been observed over a wide frequency 
range from few kHz up to several GHz. MI is related to the effective 
permeability of the sample, and hence, all mechanisms affecting 
magnetization processes of the material ought to be considered. Albeit this 
effect is usually weak, a giant magnetoimpedance effect (GMI) in amorphous 
ferromagnetic FeCoSiB wires with small magnetic fields (a few Oersteds) and 
at low frequencies (kHz to MHz) was discovered \cite{beach}. A smaller MI effect has 
also been observed, later on, in ribbons and thin magnetic films \cite{panina}.

Several reasons explain that interest. Firstly, potential applications of 
GMI in magnetic fields sensors and magnetic recording heads have been 
studied and tested in different systems (thin films, sandwich structures, 
amorphous ribbons and microwires,...). Secondly, from the theoretical 
viewpoint, a better understanding of the mechanisms that drive MI and GMI 
effects provides an additional tool to investigate intrinsic and extrinsic 
magnetic properties of soft ferromagnetic materials.

In both cases thin films and ribbons possess several advantages with respect 
to wires because they allow several orientations of $H_{dc}$ versus \textit{ac} current 
direction. Moreover, sputtered or otherwise produced films allow 
multilayering and size reduction required for integrated devices.

At frequencies above a few tens of MHz, sample length is no longer 
negligible with respect to the \textit{ac} signal wavelength and consequently, 
transmission line theory must be used \cite{brunetti}.

In this paper, an appropriate broadband frequency method for the 
determination of the MI effect, in thin films and ribbons, is presented. The 
method is based on the automated measurement, by a network analyser, of the 
reflection coefficient of a cell loaded by the film under test. The field 
$H_{dc}$ can be applied either in the plane of the sample or out of it with 
sets of Helmholtz coils. Some illustrative results obtained using this 
measurement method with a permalloy multilayer thin film are presented and 
discussed.

The samples used in the present work are single NiFe magnetic thin films of 
different thicknesses (from a few hundred to a few thousand {\AA}), NiFe 
multilayer and sandwich structures deposited by RF diode sputtering on 
20x2mm glass subtrates. These samples present an in-plane magnetization 
perpendicular or parallel to the sample axis.

The measuring device is a matched clip fixture, connected to a HP8753 
network analyser yielding the complex reflection scattering coefficient 
S$_{11}$ of the clip fixture loaded by the sample. 

The complex impedance $Z(H, f)$ of the sample, is given by

\begin{equation}
Z(H,f) = Z_0 \frac{ S_{11} (H,f) + 1 }{S_{11}(H,f) - 1}
\end{equation}

with $Z_{0}$ = 50 $\Omega $ the device characteristic impedance.

The input RF power is 0 dBm and the dc external magnetic field H, applied by 
two pairs of Helmholtz coils, can be swept from --200 Oe to 200 Oe along or 
perpendicular to the sample axis. 

The apparatus, briefly described here, allows the exploration of the MI 
effect over a broad frequency band from 0.3 MHz to 500 MHz. The lower bound 
is imposed by the analyser capabilities and the upper one by our cell 
design, nevertheless, the method may be easily extended, to lower or to 
higher frequencies.

The MI ratio used is given by the expression:

\begin{equation}
|\frac{\Delta Z}{Z}| = |{\frac{Z(H,f) - Z(H_{\max } ,f)}{Z(H_{\max },f)}}|
\end{equation}

where $H$ is the \textit{dc} applied magnetic field, $H_{max}$ the maximum value of $H$ and 
$f$ the frequency of the driving \textit{ac} current.

Numerous samples have been tested using this apparatus. As an example, the 
results obtained in the longitudinal configuration for a multilayer thin 
film is shown in the figure below. The sample is a trilayer thin film 
(Ni$_{80}$Fe$_{20}$/SiO$_{2}$/Ni$_{80}$Fe$_{20})$. The thicknesses are 960 
{\AA} for the Ni$_{80}$Fe$_{20}$ layers and 14 {\AA} for the SiO$_{2}$ 
layer. This sample exhibits an in-plane magnetization perpendicular to the 
sample longitudinal geometrical axis. The measurements are done in the 0.3 
MHz to 400 MHz frequency range and for a \textit{dc} magnetic field varying from 
-40 Oe to 40 Oe. This sandwich structure is peculiar due to the presence of 
a very thin insulating layer of SiO$_{2}$ that provides a significant 
lowering of the coercive field of the structure as we have previously shown 
\cite{gieral}. A weak coercive field is required in some MI based sensors and might be 
designed along the lines previously described in \cite{gieral}.

\begin{figure}[h!] 
\centering
\includegraphics[angle=-90, width=13.5 cm]{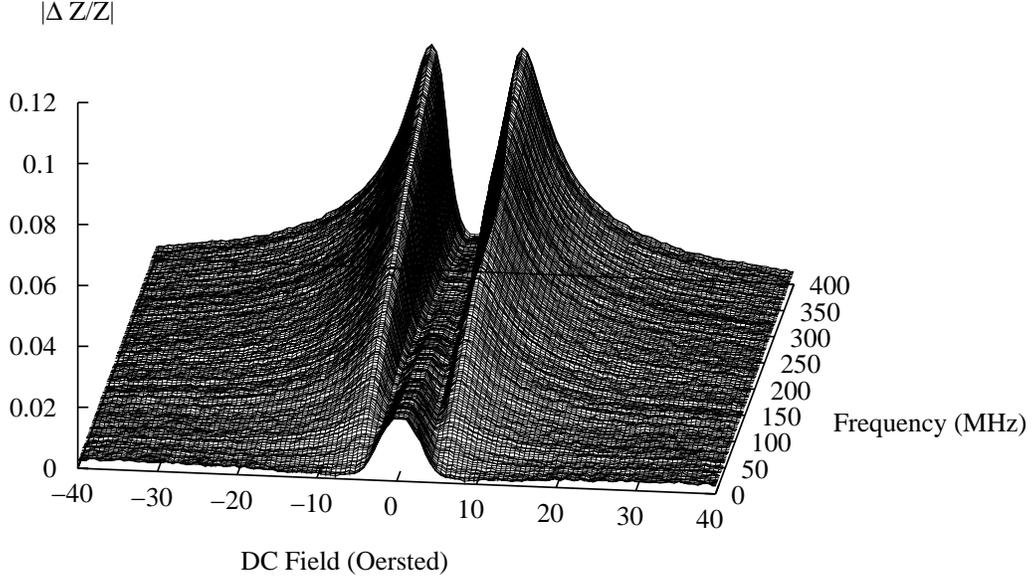} 
\caption{Three-dimensional plot of the normalised MI ratio $|\Delta Z/Z|$ for a 
Ni$_{80}$Fe$_{20}$/SiO$_{2}$/Ni$_{80}$Fe$_{20}$ sandwich structure versus frequency and field.} 
\end{figure}

For relatively low frequencies, the $|\Delta Z/Z|$ ratio decreases first quickly with 
increase of the \textit{dc} field $H$. When the frequency increases, the field dependence 
of $|\Delta Z/Z|$ gradually changes. The magnitude of the MI ratio increases with the 
increase of \textit{dc} magnetic field $H$, reaching a sharp peak for $H$ nearly equal to 
$H_{k}$, the effective anisotropy field of the sample ($ H_{k} \sim $ 4.8 Oe), and then 
gradually decreases to zero with further increase of the 
magnetic field.

\section{Conclusion}

A novel automated broadband frequency method for the determination of the MI 
effect in thin films and ribbons, based on the measurement of the reflection 
coefficient of a loaded cell, is presented. The measurement configuration, 
allows in-plane and out of plane measurements. The results obtained with 
this measurement method on permalloy thin films sandwich structures and 
CoFeSiB ribbons are in good agreement with those obtained by other 
conventional methods.


\begin{thebibliography}{00}

 \bibitem{beach} R. S. Beach and A. E. Berkowitz, ``Giant magnetic field dependent 
impedance of amorphous FeCoSiB wire'', App. Phys. Lett, vol.64, 
pp.3652-3654, 1994.

 \bibitem{panina} L. V. Panina, K. Mohri, ``Giant magneto-impedance in Co-rich amorphous 
wires and films'', IEEE trans. Magn., vol. 31, pp. 1249-1260, 1995.

 \bibitem{brunetti} L. Brunetti, P. Tiberto, F. Vinai `` A new measurement method for 
magneto-impedance in soft amorphous ferromagnets at microwave frequencies'', 
Sens, Actu., vol. A67, pp. 84-88, 1998.

 \bibitem{gieral} J. Gieraltowski and C. Tannous, ``Magnetic and transport properties of 
Ni$_{80}$Fe$_{20}$/SiO$_{2}$/Ni$_{80}$Fe$_{20}$ trilayers'' IEEE trans. 
Magn., vol. 38, pp. 2679-2681, 2002

\end{thebibliography}
\end{document}